\documentclass[fleqn,usenatbib]{mnras}

\usepackage{newtxtext,newtxmath}

\usepackage[T1]{fontenc}

\DeclareRobustCommand{\VAN}[3]{#2}
\let\VANthebibliography\thebibliography
\def\thebibliography{\DeclareRobustCommand{\VAN}[3]{##3}\VANthebibliography}

\usepackage{graphicx}	
\usepackage{amsmath}	
\usepackage{amssymb}	

\title[X-ray Pulsars Polarisation II]{Polarisation of Accreting X-ray Pulsars. II. Hercules X-1}

\author[I. Caiazzo \& J. Heyl]{
Ilaria Caiazzo,$^{1,2}$\thanks{E-mail: ilariac@caltech.edu}
Jeremy Heyl,$^{2}$
\\
$^{1}$TAPIR, Walter Burke Institute for Theoretical Physics, Mail Code 350-17, Caltech, Pasadena, CA 91125, USA\\
$^{2}$Department of Physics and Astronomy, University of British Columbia, Vancouver, BC V6T 1Z1, Canada
}

\date{Accepted XXX. Received YYY; in original form ZZZ}

\pubyear{2020}

\begin{document}
\label{firstpage}
\pagerange{\pageref{firstpage}--\pageref{lastpage}}
\maketitle

\begin{abstract}
We employ our new model for the polarised emission of accreting X-ray pulsars to describe the emission from the luminous X-ray pulsar Hercules X-1. In contrast with previous works, our model predicts the polarisation parameters independently of spectral formation, and considers the structure and dynamics of the accretion column, as well as the additional effects on propagation due to general relativity and quantum electrodynamics. We find that our model can describe the observed pulse fraction and the pulse shape of the main peak, as well as the modulation of the cyclotron line with phase. We pick two geometries, assuming a single accretion column or two columns at the magnetic poles, that can describe current observations of pulse shape and cyclotron modulation with phase. Both models predict a high polarisation fraction, between 60 and 80\% in the $1-10$~keV range, that is phase- and energy-dependent, and that peaks at the same phase as the intensity. The phase and energy dependence of the polarisation fraction and of the polarisation angle can help discern between the different geometries.
\end{abstract}




\section{Introduction}
\label{sec:intro}
The polarisation of light provides two observables, polarisation degree and angle, that are very sensitive to the anisotropies of the emission region; therefore, polarisation, combined with spectral and timing analysis, can be a powerful tool to study the geometry of astronomical sources. Regular astrophysical observations of X-ray polarisation will soon become a reality, as several observatories with an X-ray polarimeter on board are now being developed. The NASA SMEX mission \textit{IXPE} \citep{2016SPIE.9905E..17W}, in the 1--10~keV energy range, and the Indian \textit{POLIX} (5--30 keV), are both scheduled to launch in 2021, as well as the rocket-based \textit{REDSox} (0.2--0.8 keV) \citep{SPIE_REDSoX}, while the balloon-borne \textit{X-Calibur}~\citep{2014JAI.....340008B} and \textit{PoGO+} \citep{2018MNRAS.tmpL..30C} are already flying. Additionally, the Chinese--European \textit{eXTP} \citep{2016SPIE.9905E..1QZ} (1--10~keV), is scheduled for launch in 2025, while the narrow band (250 eV) \textit{LAMP}~\citep{2015SPIE.9601E..0IS} and the broad band (0.2--60~keV)  \textit{XPP} \citep{2019arXiv190409313K,2019arXiv190710190J} are still at the concept stage. Many of these upcoming polarimeters employ the gas pixel detector technology, or GPD, which was recently put to test by the CubeSat \textit{PolarLight} \citep{2019ExA....47..225F}, a small polarimeter without optics, that was able to measure the polarisation of the Crab nebula \citep{2020NatAs.tmp..100F}.

As always when a new window opens in astronomy, it is fair to expect that the new polarimetric capabilities of the upcoming telescopes in the X-rays will lead to new breakthroughs in the study of X-ray sources, in particular for accreting neutron stars and black holes. In order to reap the most from these future observations, we need to be ready with a suite of theoretical models that can constrain, when compared with observations, what is still unknown about these objects. In this paper, which is the second in a series, we present a method for modelling the polarised emission of accreting X-ray pulsars that agrees with current spectroscopic observations and that makes clear predictions of the polarisation parameters depending on the geometry of the accretion region and on the strength of the magnetic field. In the previous paper (Caiazzo \& Heyl 2020a, hereafter Paper I) we described in detail the model, whereas here we apply our method to model the polarised emission from the accreting X-ray pulsar Hercules X-1 (Her X-1).

Accreting X-ray pulsars are neutron stars that live in a close binary and accrete material from a companion star. They usually have high magnetic fields, of the order of $10^{12}-10^{13}$~G. For this reason, the accretion disc is truncated at the magnetospheric radius, several thousands kilometres from the neutron star, where the magnetic pressure from the star's dipolar magnetic field and the ram pressure in the disc become comparable, and the ionised gas is then funneled to the magnetic poles of the star along the magnetic field lines. Depending on the luminosity of the radiation at the magnetic poles, the ionised gas may be able reach the surface of the neutron star, heating the magnetic pole and creating hot-spots, or the escaping photons may be able to slow down the in-flowing gas before it reaches the surface, and therefore an accretion column may form above the pole, held together by magnetic pressure. In both cases, the small area of the emission region coupled with the rotation of the star gives origin to the pulsating nature of the emission.

Her X-1 was one of the first accreting X-ray pulsars to be discovered \citep{1972ApJ...174L.143T} and is one of the brightest and most studied; for this reason, it will be one of the main targets of all the upcoming polarimeters. Her X-1 is a persistent source in the X-rays, and its X-ray flux shows several periodic modulations. The neutron star is orbiting a 2.2~M$_\odot$ stellar companion \citep{1997MNRAS.288...43R}, HZ Her, with a period of 1.7 days, as can be inferred from the deep eclipses in the X-ray flux, while the spin period of the neutron star itself is 1.24~s. A third, super-orbital modulation, thought to be caused by the precession of the accretion disc, has a period of 35 days \citep{1973ApJ...184..227G}. The 35 days cycle shows two ``on'' states, in which the pulsed emission is observed: a ``Main-On'' of about 10 days and a 5-days ``Short-On'', separated by two ``off'' states during which the emission from the neutron star is occulted by the disc \citep{1976ApJ...209..562G,1980ApJ...237..169B,1999ApJ...510..974S,2000ApJ...539..392S,2008A&A...482..907K,2013A&A...550A.111V}.   

Intriguingly, the pulse profile evolves following the same 35-day cycle, with features appearing and disappearing with phase and energy \citep{1986ApJ...300L..63T,1998ApJ...502..802D,2000ApJ...539..392S,2013A&A...550A.110S}. This modulation in the pulse shape has been explained by the precession of the neutron star \citep{1986ApJ...300L..63T,2013MNRAS.435.1147P} or of the disc \citep{2000ApJ...539..392S,2002MNRAS.334..847L}. Both theories present some issues. The neutron star precession model implies that a very strong and not yet identified feedback mechanism links the clock of the disc's and the neutron star's precession, as the two appear perfectly synchronised on very short timescales; additionally, a mechanism is needed to explain why the precession period of the neutron star changes every few years by a few percent. On the other hand, if the appearing and disappearing of features are to be explained by the occultations of a precessing disc, the disc has to be very close to the star, at about 20-40 neutron star radii, which is more than ten times closer than the magnetospheric radius for a magnetic field of a few $10^{12}$~G. For a detailed discussion, see \citet{2013A&A...550A.110S}. Here we suggest an alternative explanation, in which the features in the pulse profile that show a modulation in the 35 days period are caused by reflections off the precessing disc of the beamed radiation coming from the neutron star (see \S~\ref{sec:pulse}).

The modulations of spectral features can provide further insights on the geometry of the source and on the origin of the emission.  The spectra of many accreting X-ray pulsars show cyclotron resonant scattering features (CRSFs), caused by resonant scattering of photons off electrons in the high magnetic field \citep{1978ApJ...219L.105T,2000ApJ...544.1067A}. The CRSFs provide a direct measurement of the strength of the magnetic field in the emission region, and that is how the magnetic field of Her X-1 has been measured to be a few times $10^{12}$~G. The cyclotron feature's energy is often modulated with rotational phase, and this has been explained by the fact that as the star rotates, the emission that we see comes from different emitting spots, where the magnetic field has different strengths \citep[e.g. in][]{2004A&A...427..975K}. In some pulsars, the CRSF shows a long term evolution as well that is thought to be due to changes in the geometry of the accretion region or of the local magnetic configuration \citep{2012A&A...544A.123B,2014A&A...572A.119S,2014ApJ...781...30N}. In particular, the centroid energy of the CRSF in Her X-1 has shown a steady decline until the year 2010 and then it has remained stable \citep[][and references therein]{2019MNRAS.484.3797J}. Some sources show an additional long term variability in the cyclotron line scattering feature that correlates (or anti-correlates) with accretion luminosity \citep{2006MNRAS.371...19T}. As the pulsars that show a negative correlation are also the brightest ones, it has been suggested that a negative correlation is also an indication of super-critical luminosity and therefore the difference in variability could be caused by a difference in accretion geometry \citep{2015MNRAS.454.2714M,2013ApJ...777..115P,2015MNRAS.448.2175L,2018MNRAS.474.5425M}.
   
Another spectral feature that is also seen varying with the pulse phase is the iron K-$\alpha$ emission line at about 6.4~keV, which is thought to be caused by fluorescence in cool gas illuminated by an X-ray beam. The location of this material is not certain, as it can be the accretion disc, a corona above it or even the atmosphere of the companion star \citep{1994ApJ...437..449C}, but the inner edge of the accretion disc seems to be the the most likely location. Using \textit{RXTE} PCA data on a ``Main-On'' state, \citet{2013A&A...550A.111V} presented a deep analysis of the evolution of spectral parameters with spin phase and with precession phase. Their findings were later confirmed by a study performed with \textit{NuSTAR} data by \citet{2013arXiv1309.5361F}.

We here apply the method developed in Paper I to reproduce many of the observational features of Her X-1 and to make a prediction of the polarisation signal. We find that in the context of our model, the modulation of the cyclotron energy with phase is naturally explained by the fact that the emission from the column is relativistically beamed, and that different viewing angles onto the column and different parts of the column (where the gas has different velocities) dominate the emission in different phases \citep{falkner2018}. Several geometries and both the one-column and the two-column model can reproduce current observations, but we show that the polarisation signal, especially the polarisation angle, is very sensitive to the geometry.

\section{The Model}
\label{sec:model}
The emission from X-ray pulsars is difficult to model from first principles, as the picture is complicated by the presence of a strong magnetic field, by the importance of radiation pressure in the description of the accretion hydrodynamics and by the fact that the emitting gas is flowing with a high bulk velocity, up to half of the speed of light. Several attempts have been made to calculate the spectral formation in accreting X-ray pulsars based on theoretical models \citep{1980ApJ...236..911Y,1981ApJ...251..278N,1985ApJ...298..147M,1985ApJ...299..138M,1996ApJ...457L..85K,2016PASJ...68...83K}; however, the models often oversimplify some aspects of the radiation generation and propagation, and the results do not always agree very well with the observed spectral profiles. On the other hand, the procedure of fitting the spectra with multicomponent functions of energy as power laws, blackbodies and exponential cut-offs is not easy to relate to the physical properties of the source.

The situation improved recently with the development by Becker and Wolff of a new model (hereafter the B\&W model) for the spectral formation \citep{2005ApJ...621L..45B,2005ApJ...630..465B,2007ApJ...654..435B,2016ApJ...831..194W,2017ApJ...835..129W,2017ApJ...835..130W}. They were able to derive a spectrum from the solution of a coupled radiation and hydrodynamic transport equation that fits very well the observed profiles and returns estimates of the properties of the accretion flow, as for example the optical thickness of the column, the temperature of the electrons and the size of the column itself. Even though several simplifying assumptions are made to keep the treatment analytic, the B\&W model is the current theoretical model that best fits observations, which however does not provide any prediction on polarisation.

In Paper I, we present a new method to model the polarised emission of accreting X-ray pulsars in the accretion column scenario, assuming that the structure of the column can be described as in the B\&W model. For the first time, our model takes into account the macroscopic structure and dynamics of the accretion column and the changes that happen during the propagation of the radiation toward the observer, including the effects of relativistic beaming, gravitational lensing and quantum electrodynamics. We find that the beaming and polarisation of the emission from the column are dominated by the presence of the strong magnetic field and by the high speed of the in-falling gas. Our calculations are exact and fully analytical and predict the polarisation parameters of the radiation emitted by the accretion column independently of radiation transfer. The results depend on the physical parameters of the accretion column, in particular on the radius and height of the accretion column, on the strength of the magnetic field, on the velocity profile of the gas inside the column and on the optical depth of the column. This information can be obtained by spectral fitting within the context of the B\&W model.

\begin{figure}
    \centering
    \includegraphics[width=0.8\columnwidth]{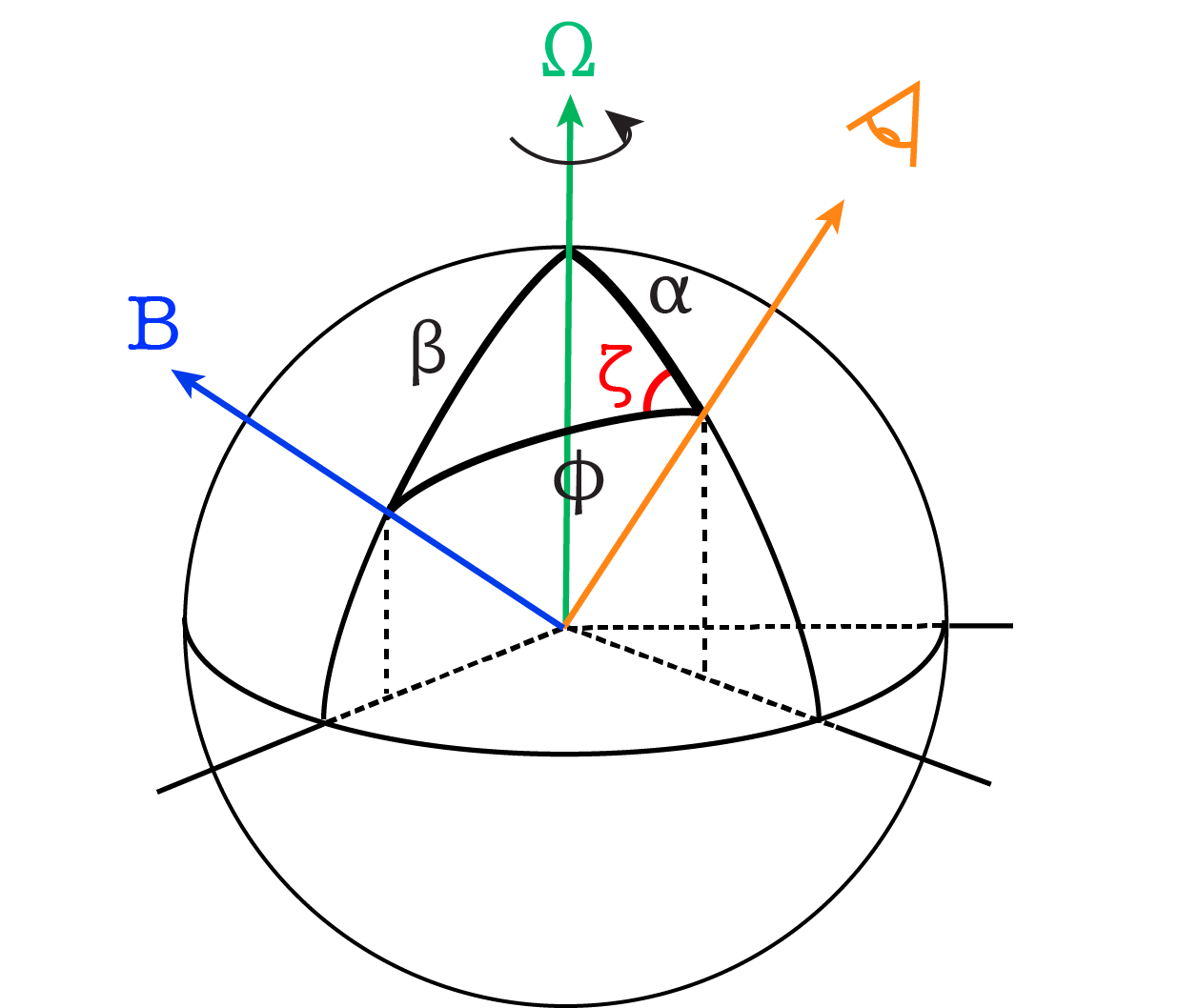}
    \caption{The geometry of the system is identified by the angle $\alpha$ between the line of sight (in orange) and the rotation axis (in green) and the angle $\beta$ between the rotation axis and the magnetic axis (in blue). The angle $\phi$ between the line of sight and the magnetic axis is the same as the phase in the case of an orthogonal rotator (see Paper I). The angle $\zeta$ (in red) between the magnetic axis and the rotation axis projected in the plane of the sky determines the polarisation angle. }
    \label{fig:Geometry}
\end{figure}

In this paper, we use the fit obtained by \citet{2016ApJ...831..194W} of the phase-averaged \textit{NuSTAR} spectrum of Her X-1. The parameters that we need for our model can be obtained from the following parameters in the paper:
\begin{itemize}
    \item the radius of the accretion column $r_0 = 107$ m;
    \item the observed cyclotron line energy $E_{\rm{cyc}} = 37.7$~keV;
    \item the dimensionless parameter $\xi= 1.36$, defined in eq.~26 of~\citep{2007ApJ...654..435B}, which determines the optical depth;
    \item the accretion rate $\dot{M}=2.59\times10^{17}$~g/s;
\end{itemize}
In addition, we employ a value of $M_*=1.5$~M$_\odot$ for the mass and of $R_*=10$~km for the radius of the neutron star, and a distance to the source of 6.6~kpc \citep{1997MNRAS.288...43R}. 
The inclination angle of Her X-1 and the angle between the rotation axis and the magnetic field axis can be somewhat constrained by analysing the pulse profile (see \S~\ref{sec:pulse}) but measuring the polarisation angle as a function of phase will put much more stringent constraints on the geometry of the source (\S~\ref{sec:pol}). Fig.~\ref{fig:Geometry} shows our convention for the names of the angles, where we have indicated the rotation axis $\hat{\mathbf{\Omega}}$ in green, the magnetic axis in in blue and the line of sight in orange. The angle between the line of sight and the rotation axis $\alpha$ and the angle between the rotation axis and the magnetic axis $\beta$ remain constant as the star rotates, while the angle between the line of sight and the magnetic field axis $\phi$ and the angle between the magnetic axis and the rotation axis projected in the plane of the sky, $\zeta$, which determines the polarisation angle, change with phase.

\section{Pulse Profile}
\label{sec:pulse}
\begin{figure}
    \centering
    \includegraphics[width=0.8\columnwidth]{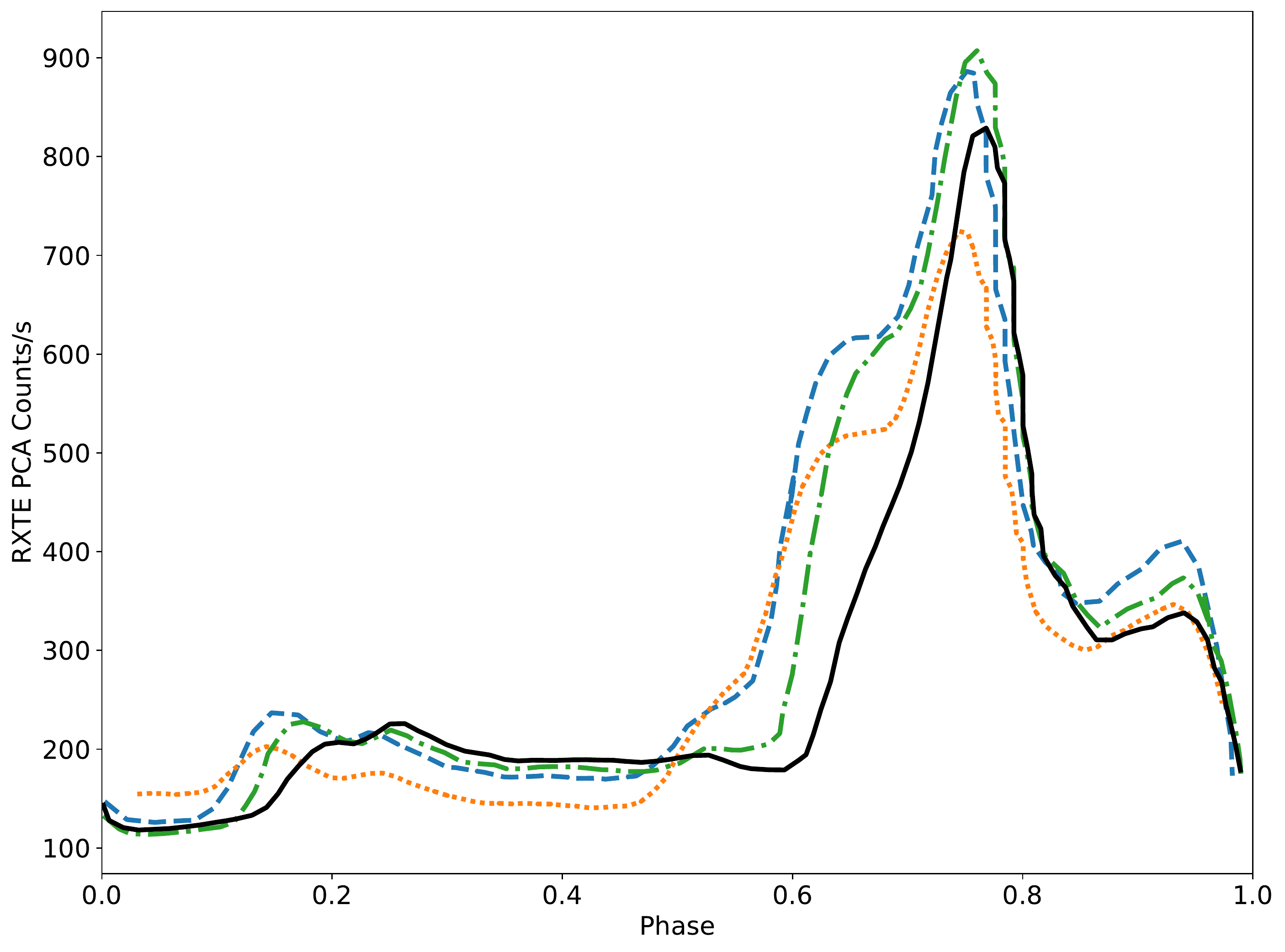}
    \caption{Pulse profiles averaged in the 9-13~keV range from \citet{2013A&A...550A.111V}. They corresponds to different phases in a ``Main On'' of the 35-day period: 0.03 (orange dotted line), 0.10 (blue dashed line), 0.15 (green dash dotted line) and 0.2 (black solid line). The uncertainties are smaller than the line width.}
    \label{fig:pulseVasco}
\end{figure}

\citet{2013A&A...550A.111V} analysed \textit{RXTE} observations of a ``Main On'' state of Her X-1, and they report the variation of spectral features with spin phase and of the pulse profile for four intervals in the precession period, those corresponding to phase 0.03, 0.10, 0.15 and 0.2 of the 35-day period. In Fig.~\ref{fig:pulseVasco} we show the pulse profiles observed by \citet[][their Fig.~2]{2013A&A...550A.111V} averaged over the energy interval 9-13~keV. The main peak at about $\sim0.75$ in pulse phase is surrounded by two ``shoulders'': a left shoulder that shows strong variations with precession phase and that is almost absent in precession phase 0.2 (black solid line) and a right shoulder that only varies as much as the main peak with precession phase. \citet{2013A&A...550A.111V} also report the variation of the intensity of the iron K-$\alpha$ line with pulse phase and precession phase: the intensity is quite constant with phase except in the peak, where it is undetectable, and in the left shoulder, where it is higher than average. Additionally, in the phases of the precession cycle where the left shoulder is less pronounced (like in phase 0.2), the peak in the iron line at the shoulder location is also less pronounced. 

The changes in intensity of the main peak during the ``Main-On'' are associated with the precession of the accretion disc: in the first phases, the disc gradually clears the view to the the X-ray emitting region, and the flux increases, while toward the end of the ``Main-On'', the inner edge of the disc starts blocking the emission region and the flux declines. The modulation in the left shoulder is much more pronounced than the modulation in the peak, and it does not follow the same trend. Moreover, while the iron line is undetectable at the main peak, in the left shoulder the line reaches its maximum intensity and it shares the same modulation with the left shoulder over the 35-day precession period. As the iron fluorescence is generated when an X-ray beam is reprocessed by cool material, this trend hints strongly that the left shoulder is coming from radiation reflected by the precessing disc. The left shoulder comes immediately before the main peak in time, which would suggest that it is caused by the reflection off the disc of the highly beamed emission from the column, pointing toward the disc right before pointing toward the observer. The right shoulder does not show a strong modulation with precession phase, and the strength of the iron line remains close to the average value in the phase of the right shoulder. Therefore, there is no particular indication whether the right shoulder is the result of a reflection off the disc or off the neutron star surface. It could be either or a combination of both.

\begin{figure}
    \centering
    \includegraphics[width=\columnwidth]{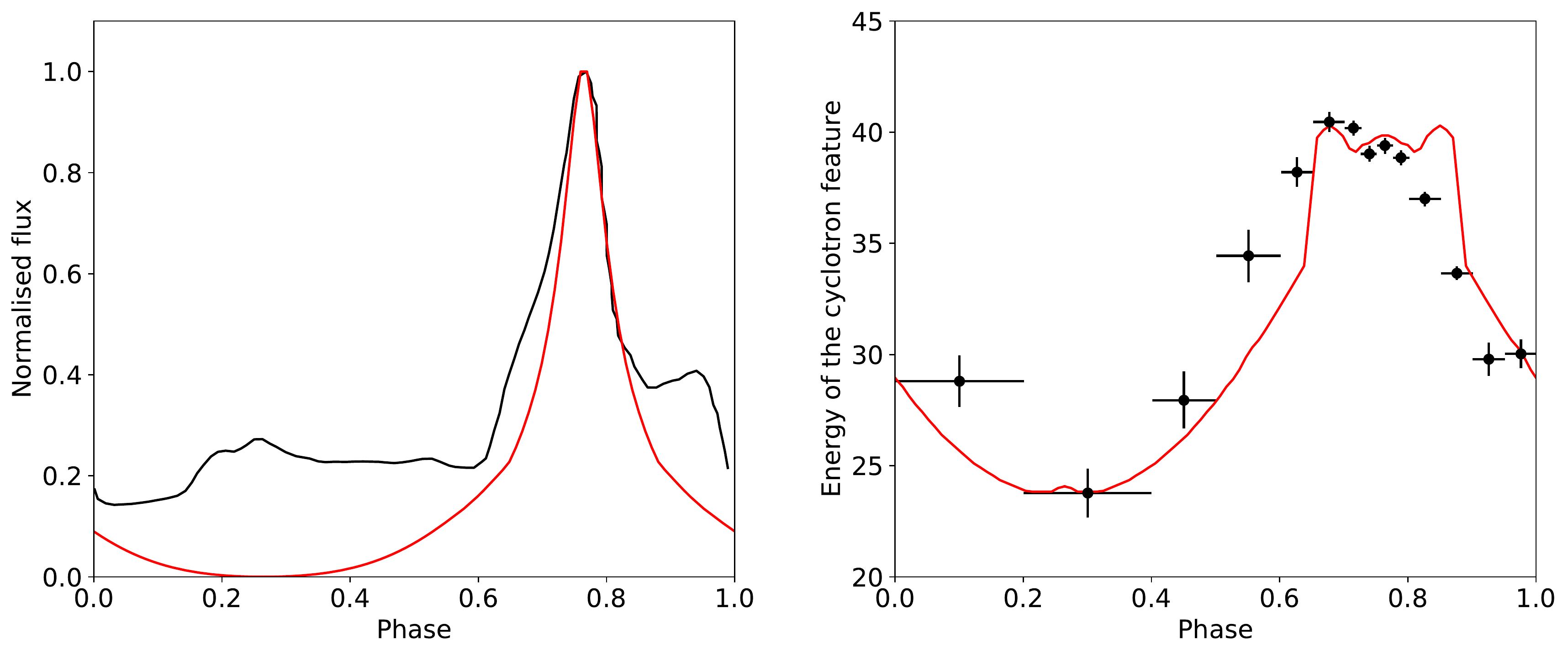}
    \includegraphics[width=\columnwidth]{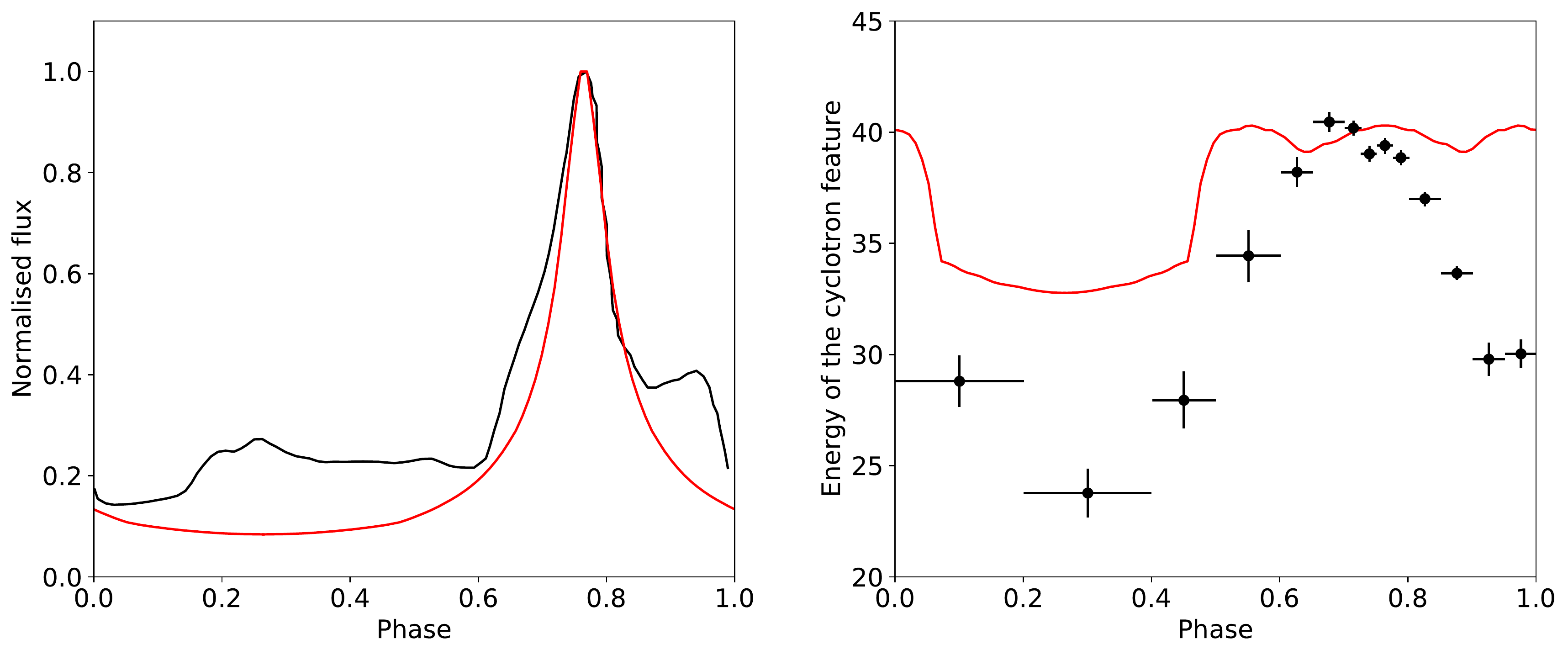}
    \caption{In the left panels, in black, the observed pulse profile for Her X-1 is shown for the phase in the precession cycle with the least contribution from the disc (phase 0.2, same as black solid line in Fig.~\ref{fig:pulseVasco}) while in the right panels, the weighted mean values for the phase-dependent cyclotron line energy averaged over the 35-day precession period are plotted in black with error bars \citep[data from][]{2013A&A...550A.111V}. The solid red line shows the prediction for the one-column model for different geometries: in the top panels, $\alpha=83^\circ$ and $\beta=86^\circ$, while in the lower panels $\alpha=30^\circ$ and $\beta=155^\circ$.}
    \label{fig:pulseOne}
\end{figure}

\begin{figure}
    \centering
    \includegraphics[width=\columnwidth]{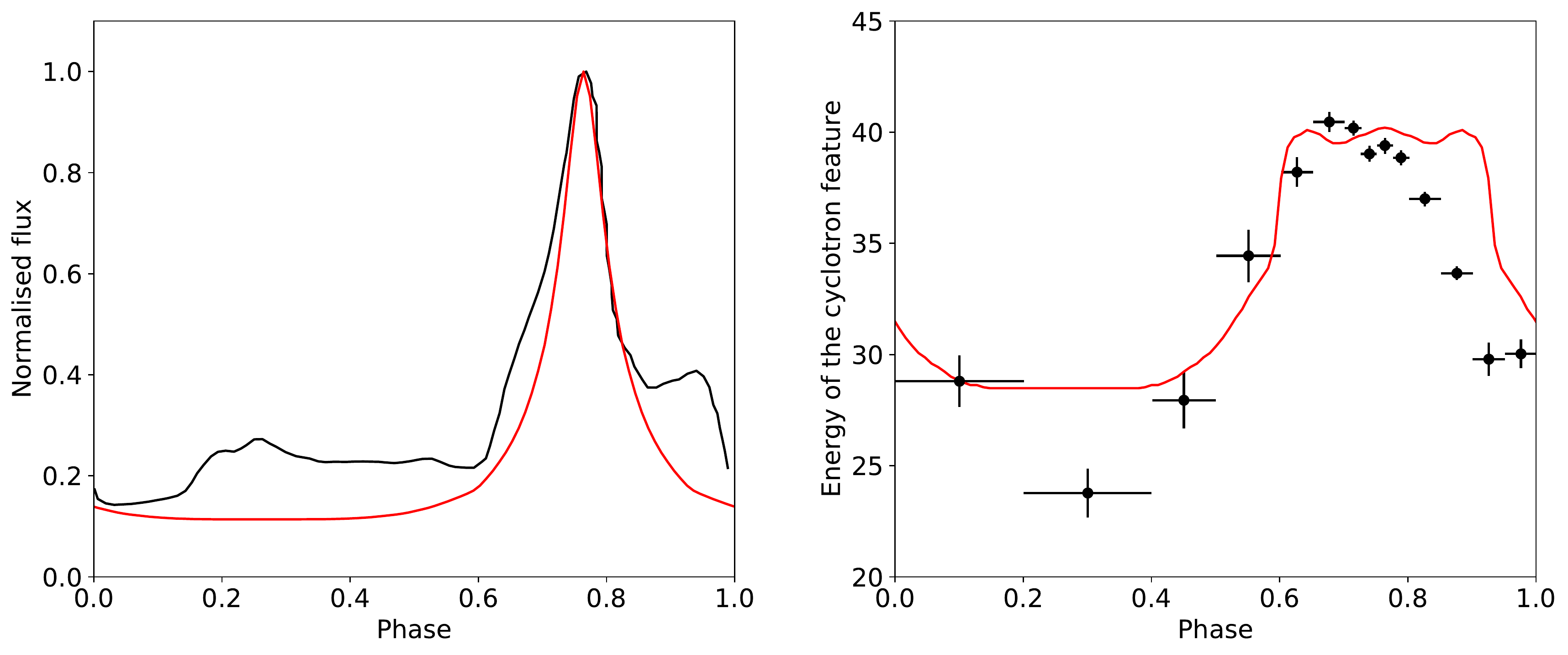}
    \includegraphics[width=\columnwidth]{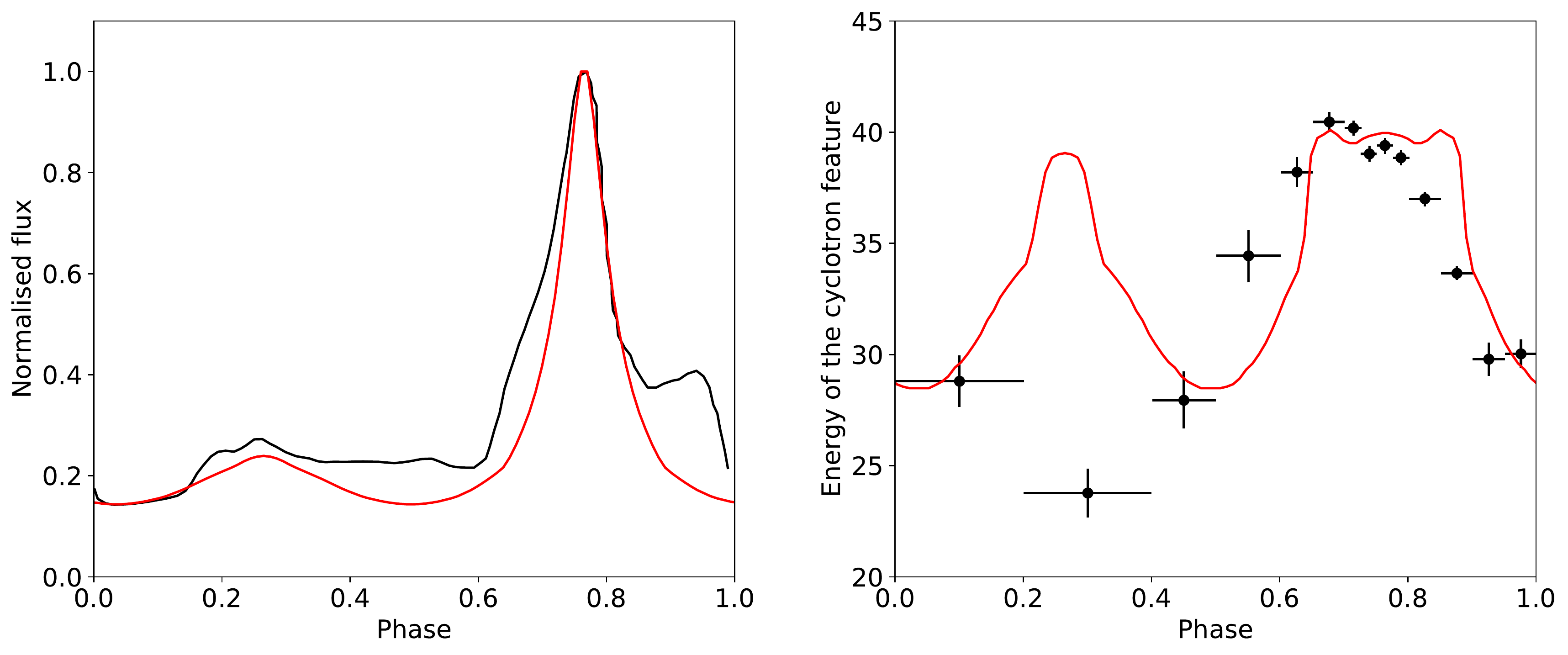}
    \caption{Same as in Fig.~\ref{fig:pulseOne} but for the two-column model and different geometries: in the top panels, $\alpha=50^\circ$ and $\beta=42^\circ$, while in the lower panels $\alpha=75^\circ$ and $\beta=115^\circ$.}
    \label{fig:pulseTwo}
\end{figure}

We here assume that at the main peak, the emission from the accretion column, which is highly beamed, is pointing toward the observer, and that the varying emission in the phases outside of the main peak comes from reflections off the disc and/or the neutron star surface. We will discuss the effect on polarisation of the two options in \S~\ref{sec:pol}. Additionally, the disc direct emission could contribute, but we do not expect it to change with rotation phase and therefore it could only provide a constant background. Within our model, we can reproduce the pulse fraction and the shape of the peak both for the case in which the X-rays are emitted by two accretion columns at the magnetic poles of the neutron star or by a single column. There are several geometries, identified by the angles $\alpha$ and $\beta$ in Fig.~\ref{fig:Geometry}, that produce a pulse profile that agrees with the observed one. The left panels of Figs.~\ref{fig:pulseOne} and~\ref{fig:pulseTwo} show only some examples. In the left panels of Fig.~\ref{fig:pulseOne}, the predicted pulse profile from the one-column model is shown in red for $\alpha=83^\circ$ and $\beta=86^\circ$ (upper panel) and for $\alpha=30^\circ$ and $\beta=155^\circ$ (lower panel), while in black we can see the pulse profile for the precession phase where the left shoulder is less pronounced (phase 0.2, also in black in Fig.~\ref{fig:pulseVasco}). In the left panels of Fig.~\ref{fig:pulseTwo}, the predicted pulse profile from the two-column model is shown for $\alpha=50^\circ$ and $\beta=42^\circ$ (upper panel) and for $\alpha=75^\circ$ and $\beta=115^\circ$ (lower panel). We can see that in both figures, the different geometries can reproduce the pulse shape at the main peak, and the main difference between the geometries is how much the accretion column(s) contribute to the emission during the phases outside of the main peak.

\section{Cyclotron feature}
\citet{2013A&A...550A.111V} and \citet{2013arXiv1309.5361F} analysed the evolution of the centroid energy of the CRSF in Her X-1 using data from \textit{RXTE} and \textit{NuSTAR}. They both found that the cyclotron line energy follows roughly the pulse profile, with an almost sinusoidal shape, and that there is no indication for a modulation over the 35-day period.
The variation in cyclotron energy with phase, which has been observed in may X-ray pulsars, has been been previously explained by the fact that as the star rotates, we see radiation coming from different hot-spots on the surface of the neutron star where the strength of the magnetic field is different \citep[e.g. in][and references therein]{2004A&A...427..975K}. 
In our model, the modulation of the line centroid with phase is naturally explained by the fact that the scattering electrons in the accretion column are highly relativistic. In fact, at the top of the column, the ionised gas is moving at about half the speed of light, and the radiation emitted by the column is highly beamed toward the surface of the star. For this reason, the main peak in the pulse profile corresponds to when one of the accretion columns is almost behind the star ($\phi$ is close to $\pi$ in Fig.~\ref{fig:Geometry}) and the strong beaming causes the cyclotron energy to be shifted to higher values. As the star rotates, we see the column at smaller angles, and therefore the beaming is less intense and the cyclotron line energy decreases. 

The right panels of Figs.~\ref{fig:pulseOne} and~\ref{fig:pulseTwo} show the weighted mean values for the phase-dependent cyclotron line energy averaged over the 35-day precession period in black with error bars, as measured by \citet[][their Fig.~5]{2013A&A...550A.111V}, together with the predictions of our model, in red, for the same geometries described in \S~\ref{sec:pulse}. In the upper right panel of Fig.~\ref{fig:pulseOne}, which is the case of an almost orthogonal rotator with one column, the modulation in cyclotron energy predicted by the model seems to best reproduce the observed one. In the lower panel of the same figure, the single column is always at a high angle $\phi$ with respect to the observer, and therefore the effect of the beaming on the CRSF energy is always important, reducing the modulation with phase. Fig.~\ref{fig:pulseTwo} shows the two-columns case, in which the contribution from the second column increases the minimum in the cyclotron line energy in the upper right panel and even produces a second peak in the lower right panel. It is worth mentioning that we have not adjusted our model in any way to meet the data, and that these predictions simply come from the value of the cyclotron line estimated in \citep{2016ApJ...831..194W} and from choosing the geometry that best reproduces the main peak in flux.

Figs.~\ref{fig:pulseOne} and~\ref{fig:pulseTwo} only show the prediction for the emission from the column in red. If part of the emission in the phases outside the main peak is coming from the  neutron star surface, for example because of reflection or reprocessing, the modulation of the cyclotron line energy can be affected by the fact that in the reflecting region the magnetic field has a different strength than in the column. For example, if the right shoulder is indeed caused by a reflection on the surface, away from the magnetic pole, the local magnetic field would be lower than in the column and this could explain why the observed cyclotron energy declines faster with phase after the main peak in the observations compared to the model prediction. For the same reason, the fact that the two-column model fails to reproduce the minimum in the cyclotron energy modulation could be due to the fact that at that phase, the emission from the column is a small contribution to the total emission and that reflections from the surface could be dominating the emission. An indication could come from the width of the cyclotron feature: when different regions where the magnetic field strength has different values contribute to the emission, we expect the cyclotron feature to be wide. \citet{2013A&A...550A.111V} could not resolve the width of the cyclotron feature on the 35-day period because of large scatter in the data, but they found that on average the cyclotron feature is wider in the phases between 0.0 and 0.4.

\begin{figure}
    \centering
    \includegraphics[width=\columnwidth]{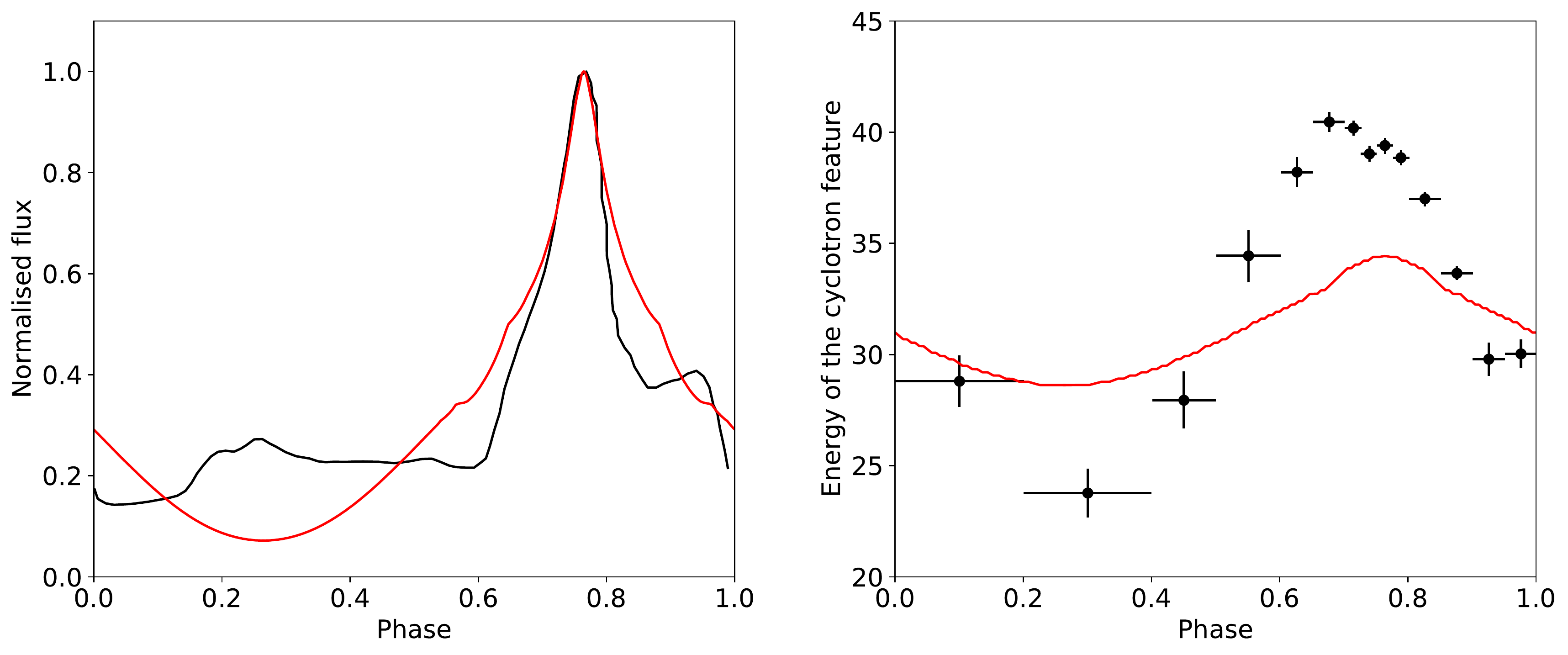}
    \includegraphics[width=\columnwidth]{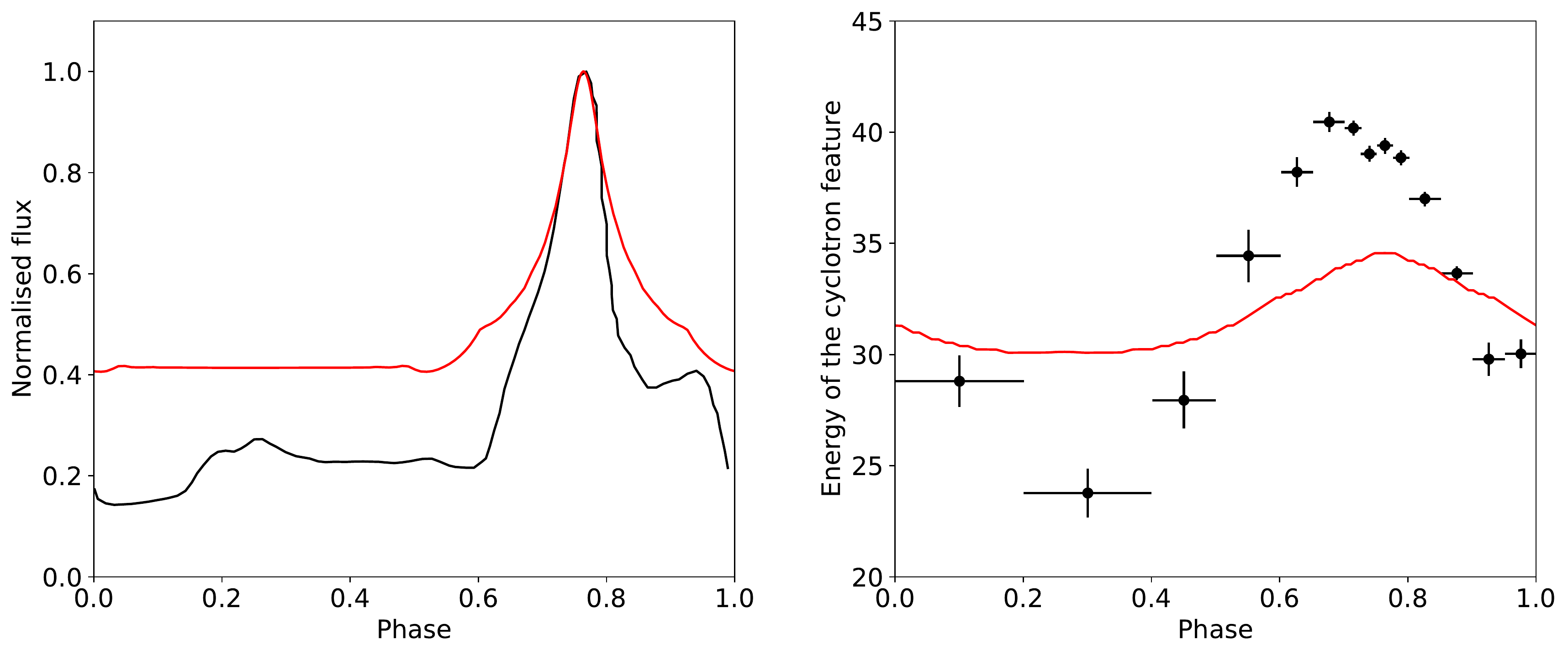}
    \caption{Same as in Fig.~\ref{fig:pulseOne} but for the short one-column model with $\alpha=67^\circ$ and $\beta=120^\circ$ (upper panels) and for the short two-columns model with $\alpha=46^\circ$ and $\beta=40^\circ$ (lower panels)}
    \label{fig:pulseShort}
\end{figure}

In Paper I, we analysed the possibility of the accretion column being shorter than what is predicted by the B\&W model. In particular, with the parameters obtained by \citet{2016ApJ...831..194W} for Her X-1, the B\&W model predicts a column height $z_{\rm{max}}=6.6$~km. If a different assumption is made on the velocity profile of the in-flowing gas and on the nature of the shock at the top of the column, the height of the column could be different. For example, in Paper I, we analysed the case in which an adiabatic shock at the top of the column drastically reduces the speed of the incoming gas, reducing also the height of the column to about 1.4~km. The predictions of the short-column model are shown in Fig.~\ref{fig:pulseShort} for the one and the two-column case in the geometries that more closely reproduce the pulse profile. In the shorter column model, the scattering electrons are moving much slower, and therefore the relativistic beaming is less pronounced. For this reason, the short-column model fails to reproduce both the narrow peak observed in the pulse profile of Her X-1 and the strong modulation of the cyclotron line energy. We therefore conclude that, even if the accretion column could in principle be shorter than what is predicted by the B\&W model, it cannot be too short, or conversely, the speed of the electrons cannot be too low, or the model fails to reproduce the observed profiles in flux and cyclotron energy.

\section{Polarisation signal}
\label{sec:pol}

\begin{figure*}
    \centering
    \includegraphics[width=\textwidth]{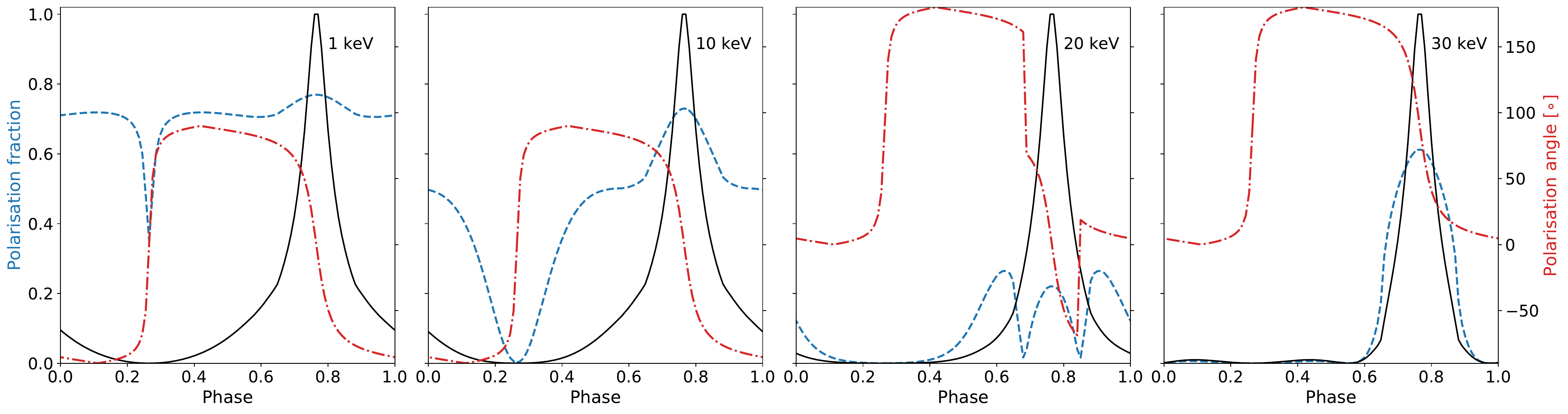}
    \includegraphics[width=\textwidth]{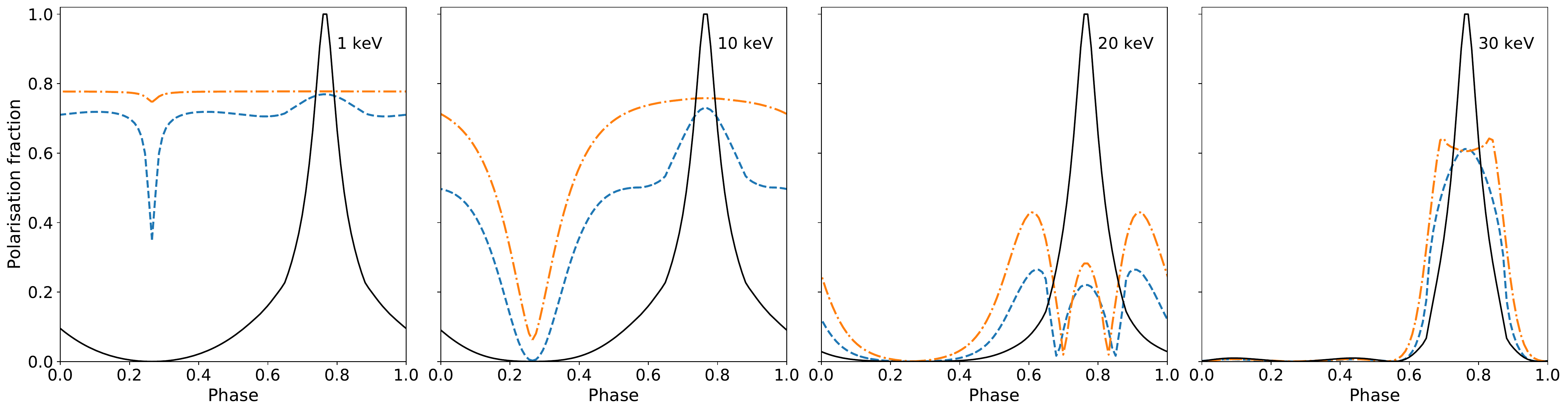}
    \caption{Polarisation parameters for the emission from the column in the one-column model with $\alpha=83^\circ$ and $\beta=86^\circ$ as function of phase and for four photon energies at the observer, from left to right: 1~keV, 10~keV, 20~keV and 30~keV. Upper panels: polarisation fraction (blue dashed line) and polarisation angle (red dash-dotted line). Lower panels: polarisation fraction including the effect of QED (blue dashed line, same as upper panels) and without QED (orange dash-dotted line)}
    \label{fig:finalpol}
\end{figure*}

\begin{figure*}
    \centering
    \includegraphics[width=\textwidth]{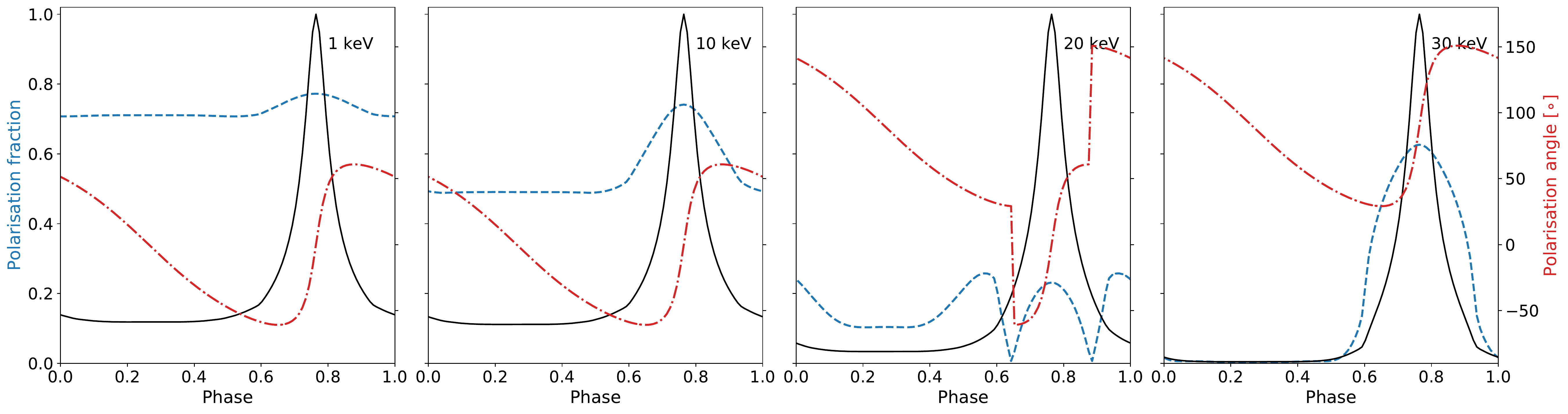}
    \includegraphics[width=\textwidth]{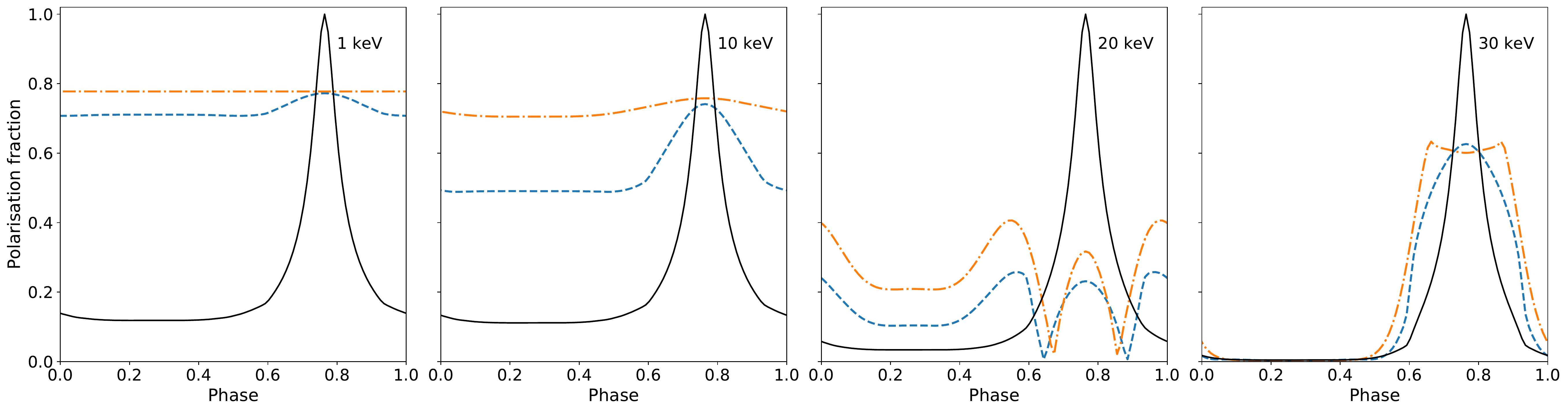}
    \caption{Same as Fig.~\ref{fig:finalpol} but for the total emission from the columns in the two-columns model with $\alpha=50^\circ$ and $\beta=42^\circ$.}
    \label{fig:finalpol2}
\end{figure*}

From the analysis of the pulse profile and of the CRSF, we see that the models that best reproduce the observed profiles are the one-column model with $\alpha=83^\circ$ and $\beta=86^\circ$ and the two-column model with $\alpha=50^\circ$ and $\beta=42^\circ$ (or equivalent geometries). We now analyse what is the polarisation signal that we expect from these models. The upper panels of Fig.~\ref{fig:finalpol} and Fig.~\ref{fig:finalpol2} show the polarisation parameters, polarisation fraction and angle, of the emission from the column(s) as a function of phase for different photon energies. In both models, the polarisation fraction (in blue) is very high at low energy and reaches a broad peak in the phase that coincides with the narrower peak in flux (in black). The main difference in polarisation fraction between the two models is that the one-column model has a deep minimum in the anti-phase of the peak, while the polarisation fraction in the two-column model is quite flat outside the peak.

As expected, the polarisation angle (in red) is quite different between the two models. In both models, below 10~keV, the light is polarised in the ordinary mode, so the polarisation direction coincides with the direction of the magnetic field in the sky. In the one-column model, the polarisation angle has a swing coincidentally with the peak because the column is going behind the star, and therefore the polarisation angle swings from $\zeta\sim90^\circ$ to $\zeta\sim-90^\circ$. The fact that $\alpha$ is close to $90^\circ$ means that we see an additional, very sharp swing at the anti-phase of the main peak. For the two-column model, in the particular geometry that we picked, the swing in the peak only goes between $\zeta\sim-60^\circ$ and $\zeta\sim60^\circ$ because $\beta$ is lower, and the swing at the anti-peak is much smoother because $\alpha$ is also lower.

In both models, the emission at 30~keV, close to the cyclotron energy, is polarised in the direction perpendicular to the magnetic axis (in the extraordinary mode) and therefore the polarisation angle follows the same pattern as at low energy, only shfted by $90^\circ$. At 20~keV, the polarisation is in the ordinary mode in the peak (around 0.75 in phase), but it goes through a zero and changes to the extraordinary mode at phases $\sim0.7$ and $\sim0.85$, which results in a shift of the polarisation angle of $90^\circ$ outside the peak. 

The lower panels of Fig.~\ref{fig:finalpol} and Fig.~\ref{fig:finalpol2} show the comparison between the polarisation fraction calculated without including the QED effect of vacuum birefringence (orange dash-and-dotted line) and including QED (blue dashed line). We can see that the effect of QED is mainly to reduce the polarisation fraction everywhere and especially outside of the peak. More details can be found in Paper I.

Figs.~\ref{fig:finalpol} and~\ref{fig:finalpol2} only depict the emission coming from the accretion column(s). As discussed in \S~\ref{sec:pulse}, outside of the peak we expect a large contribution to come from either the accretion disc or the surface of the neutron star. If the disc itself is emitting in the energy range of interest, such emission should stay constant with rotation phase and show variability with the 35-day period. A direct contribution from the disc would dilute the polarisation shown in Figs.~\ref{fig:finalpol} and~\ref{fig:finalpol2} because the emission from an accretion disc is expected to be weakly polarised, with a peak of about 11\% parallel to the disc plane if the disc is seen edge-on \citep[e.g.][and references therein]{1960ratr.book.....C,2009apj...703..569d,2009apj...701.1175s,2010apj...712..908s,2018phrvd..97h3001c}. Although direct emission from the disc would stay constant with phase, a reflection component could instead change with phase. For example, we suggested in \S~\ref{sec:pulse} that the ``left shoulder'' in the pulse profile could be caused by the disc reflecting the beamed X-ray emission from the accretion column. In this case, we expect the scattered radiation to be polarised in the same direction as the pulse and to be highly polarised. If reflected emission from the disc contributes to phases far from the main pulse, we expect it to be polarised perpendicular to the disc.

The ``right shoulder'' in the pulse profile could come from either a reflection off the disc or off the neutron star surface. As we have shown in Paper I, the radiation from the accretion column is beamed downward by special relativity and bent downward toward the stellar surface by gravity. We therefore expect a substantial portion of the surface to be illuminated by the column. Depending on whether the heat is deposited deep inside the stellar surface or if the illuminating photons are scattered in the upper layers of the atmosphere, the expected polarisation signal is quite different. In the former case, we expect the average energy of the emission in the ``right shoulder'' to be less than near the peak, the polarisation to be mainly in the X-mode for low energies and the polarisation fraction to decrease with energy \cite[see][and references therein]{2019ASSL..460..301C}. In the latter case, the illuminating photons are scattered and possibly boosted in energy through Comptonisation off the surface and therefore we expect the average energy of the emission to be higher than or similar to the one near the peak. The polarisation signal would be also very different, as the scattered photons would be polarised mainly in the O-mode \citep[see][and Paper I]{1985ApJ...299..138M}.

If the neutron star is precessing with a 35-day period, as suggested by some authors \citep{1972Natur.239..325B,1986ApJ...300L..63T,2013MNRAS.435.1147P}, we expect the polarisation direction both for the main emission and the reflected emission to shift with the precession period.

\section{Conclusions}

We have presented a model for the polarised emission from one or two accretion columns in the X-ray pulsar Her X-1. This model is able to reproduce the pulse fraction and the pulse shape of the main peak in the pulse profile of Her X-1. Additionally, the observed modulation with rotational phase of the cyclotron energy is naturally reproduced by some geometries in our model that also reproduce the pulse profile.  The model is built upon the B\&W model (see Paper I) for the phase-averaged spectrum for Her X-1, so it also is able to satisfactorily reproduce the current observations and to build upon these to make clear predictions of the expected polarisation signal. Observations of the polarised signal will help constraining the geometry of the system, in particular the nature of the 35-day super-orbital period, by determining if the disc or the neutron star precess, or both. Furthermore, polarisation will help decide the nature of the emission outside the main pulse that still varies periodically with pulsar period, specifically if it is reprocessed by the disc or the stellar surface, and in the latter case, how deeply the energy is deposited in the neutron star's atmosphere. The dawn of astrophysical X-ray polarimetry will dramatically enhance our understanding of X-ray pulsars in general and Her X-1 in particular.

\section*{Acknowledgements}

We would like to thank Michael Wolff, Peter Becker and the XMAG collaboration for for valuable input, and Sterl Phinney for useful comments. The research was supported by NSERC Canada, Compute Canada, a Burke Fellowship at Caltech and a Four-Year Fellowship at UBC.




\bibliographystyle{mnras}
\bibliography{main} 

\begin{thebibliography}{}
\makeatletter
\relax
\def\mn@urlcharsother{\let\do\@makeother \do\$\do\&\do\#\do\^\do\_\do\%\do\~}
\def\mn@doi{\begingroup\mn@urlcharsother \@ifnextchar [ {\mn@doi@}
  {\mn@doi@[]}}
\def\mn@doi@[#1]#2{\def\@tempa{#1}\ifx\@tempa\@empty \href
  {http://dx.doi.org/#2} {doi:#2}\else \href {http://dx.doi.org/#2} {#1}\fi
  \endgroup}
\def\mn@eprint#1#2{\mn@eprint@#1:#2::\@nil}
\def\mn@eprint@arXiv#1{\href {http://arxiv.org/abs/#1} {{\tt arXiv:#1}}}
\def\mn@eprint@dblp#1{\href {http://dblp.uni-trier.de/rec/bibtex/#1.xml}
  {dblp:#1}}
\def\mn@eprint@#1:#2:#3:#4\@nil{\def\@tempa {#1}\def\@tempb {#2}\def\@tempc
  {#3}\ifx \@tempc \@empty \let \@tempc \@tempb \let \@tempb \@tempa \fi \ifx
  \@tempb \@empty \def\@tempb {arXiv}\fi \@ifundefined
  {mn@eprint@\@tempb}{\@tempb:\@tempc}{\expandafter \expandafter \csname
  mn@eprint@\@tempb\endcsname \expandafter{\@tempc}}}

\bibitem[\protect\citeauthoryear{{Araya-G{\'o}chez} \&
  {Harding}}{{Araya-G{\'o}chez} \& {Harding}}{2000}]{2000ApJ...544.1067A}
{Araya-G{\'o}chez} R.~A.,  {Harding} A.~K.,  2000, \mn@doi [\apj]
  {10.1086/317224}, \href
  {https://ui.adsabs.harvard.edu/abs/2000ApJ...544.1067A} {544, 1067}

\bibitem[\protect\citeauthoryear{{Becker} \& {Wolff}}{{Becker} \&
  {Wolff}}{2005a}]{2005ApJ...621L..45B}
{Becker} P.~A.,  {Wolff} M.~T.,  2005a, \mn@doi [\apjl] {10.1086/428927}, \href
  {https://ui.adsabs.harvard.edu/abs/2005ApJ...621L..45B} {621, L45}

\bibitem[\protect\citeauthoryear{{Becker} \& {Wolff}}{{Becker} \&
  {Wolff}}{2005b}]{2005ApJ...630..465B}
{Becker} P.~A.,  {Wolff} M.~T.,  2005b, \mn@doi [\apj] {10.1086/431720}, \href
  {https://ui.adsabs.harvard.edu/abs/2005ApJ...630..465B} {630, 465}

\bibitem[\protect\citeauthoryear{{Becker} \& {Wolff}}{{Becker} \&
  {Wolff}}{2007}]{2007ApJ...654..435B}
{Becker} P.~A.,  {Wolff} M.~T.,  2007, \mn@doi [\apj] {10.1086/509108}, \href
  {https://ui.adsabs.harvard.edu/abs/2007ApJ...654..435B} {654, 435}

\bibitem[\protect\citeauthoryear{{Becker} et~al.,}{{Becker}
  et~al.}{2012}]{2012A&A...544A.123B}
{Becker} P.~A.,  et~al., 2012, \mn@doi [\aap] {10.1051/0004-6361/201219065},
  \href {https://ui.adsabs.harvard.edu/abs/2012A&A...544A.123B} {544, A123}

\bibitem[\protect\citeauthoryear{{Beilicke} et~al.,}{{Beilicke}
  et~al.}{2014}]{2014JAI.....340008B}
{Beilicke} M.,  et~al., 2014, \mn@doi [Journal of Astronomical Instrumentation]
  {10.1142/S225117171440008X}, \href
  {http://adsabs.harvard.edu/abs/2014JAI.....340008B} {3, 1440008}

\bibitem[\protect\citeauthoryear{{Boynton}, {Crosa}  \& {Deeter}}{{Boynton}
  et~al.}{1980}]{1980ApJ...237..169B}
{Boynton} P.~E.,  {Crosa} L.~M.,   {Deeter} J.~E.,  1980, \mn@doi [\apj]
  {10.1086/157856}, \href
  {https://ui.adsabs.harvard.edu/abs/1980ApJ...237..169B} {237, 169}

\bibitem[\protect\citeauthoryear{{Brecher}}{{Brecher}}{1972}]{1972Natur.239..325B}
{Brecher} K.,  1972, \mn@doi [\nat] {10.1038/239325a0}, \href
  {https://ui.adsabs.harvard.edu/abs/1972Natur.239..325B} {239, 325}

\bibitem[\protect\citeauthoryear{{Caiazzo} \& {Heyl}}{{Caiazzo} \&
  {Heyl}}{2018}]{2018phrvd..97h3001c}
{Caiazzo} I.,  {Heyl} J.,  2018, \mn@doi [\prd] {10.1103/PhysRevD.97.083001},
  \href {http://adsabs.harvard.edu/abs/2018PhRvD..97h3001C} {97, 083001}

\bibitem[\protect\citeauthoryear{{Caiazzo}, {Heyl}  \& {Turolla}}{{Caiazzo}
  et~al.}{2019}]{2019ASSL..460..301C}
{Caiazzo} I.,  {Heyl} J.,   {Turolla} R.,  2019, {Polarimetry of Magnetars and
  Isolated Neutron Stars}.
p.~301, \mn@doi{10.1007/978-3-030-19715-5_12}

\bibitem[\protect\citeauthoryear{{Chandrasekhar}}{{Chandrasekhar}}{1960}]{1960ratr.book.....C}
{Chandrasekhar} S.,  1960, {Radiative transfer}

\bibitem[\protect\citeauthoryear{{Chauvin} et~al.,}{{Chauvin}
  et~al.}{2018}]{2018MNRAS.tmpL..30C}
{Chauvin} M.,  et~al., 2018, \mn@doi [\mnras] {10.1093/mnrasl/sly027}, \href
  {http://adsabs.harvard.edu/abs/2018MNRAS.tmpL..30C} {77, L45–L49}

\bibitem[\protect\citeauthoryear{{Choi}, {Nagase}, {Makino}, {Dotani},
  {Kitamoto}  \& {Takahama}}{{Choi} et~al.}{1994}]{1994ApJ...437..449C}
{Choi} C.~S.,  {Nagase} F.,  {Makino} F.,  {Dotani} T.,  {Kitamoto} S.,
  {Takahama} S.,  1994, \mn@doi [\apj] {10.1086/175008}, \href
  {https://ui.adsabs.harvard.edu/abs/1994ApJ...437..449C} {437, 449}

\bibitem[\protect\citeauthoryear{{Davis}, {Blaes}, {Hirose}  \&
  {Krolik}}{{Davis} et~al.}{2009}]{2009apj...703..569d}
{Davis} S.~W.,  {Blaes} O.~M.,  {Hirose} S.,   {Krolik} J.~H.,  2009, \mn@doi
  [\apj] {10.1088/0004-637X/703/1/569}, \href
  {http://adsabs.harvard.edu/abs/2009ApJ...703..569D} {703, 569}

\bibitem[\protect\citeauthoryear{{Deeter}, {Scott}, {Boynton}, {Miyamoto},
  {Kitamoto}, {Takahama}  \& {Nagase}}{{Deeter}
  et~al.}{1998}]{1998ApJ...502..802D}
{Deeter} J.~E.,  {Scott} D.~M.,  {Boynton} P.~E.,  {Miyamoto} S.,  {Kitamoto}
  S.,  {Takahama} S.,   {Nagase} F.,  1998, \mn@doi [\apj] {10.1086/305910},
  \href {https://ui.adsabs.harvard.edu/abs/1998ApJ...502..802D} {502, 802}

\bibitem[\protect\citeauthoryear{Falkner}{Falkner}{2018}]{falkner2018}
Falkner S.,  2018, PhD thesis, Friedrich-Alexander-Universit\"at
  Erlangen-N\"urnberg

\bibitem[\protect\citeauthoryear{{Feng} et~al.,}{{Feng}
  et~al.}{2019}]{2019ExA....47..225F}
{Feng} H.,  et~al., 2019, \mn@doi [Experimental Astronomy]
  {10.1007/s10686-019-09625-z}, \href
  {https://ui.adsabs.harvard.edu/abs/2019ExA....47..225F} {47, 225}

\bibitem[\protect\citeauthoryear{{Feng} et~al.,}{{Feng}
  et~al.}{2020}]{2020NatAs.tmp..100F}
{Feng} H.,  et~al., 2020, \mn@doi [Nature Astronomy]
  {10.1038/s41550-020-1088-1}, \href
  {https://ui.adsabs.harvard.edu/abs/2020NatAs.tmp..100F} {4, 511}

\bibitem[\protect\citeauthoryear{{F\"{u}rst} et~al.,}{{F\"{u}rst}
  et~al.}{2013}]{2013arXiv1309.5361F}
{F\"{u}rst} F.,  et~al., 2013, arXiv e-prints, \href
  {https://ui.adsabs.harvard.edu/abs/2013arXiv1309.5361F} {p. arXiv:1309.5361}

\bibitem[\protect\citeauthoryear{{Gaenther}, {Egan}, {Heilmann}, {Heine},
  {Hellickson}, {Frost}, {Schulz}  \& {Theriault-Shay}}{{Gaenther}
  et~al.}{2017}]{SPIE_REDSoX}
{Gaenther} H.~M.,  {Egan} M.,  {Heilmann} R.~K.,  {Heine} S. N.~T.,
  {Hellickson} T.,  {Frost} J.and~{Marshall} H.~L.,  {Schulz} N.~S.,
  {Theriault-Shay} A.,  2017. pp 10399 -- 10399 -- 13,
  \mn@doi{10.1117/12.2273772}, \url {http://dx.doi.org/10.1117/12.2273772}

\bibitem[\protect\citeauthoryear{{Gerend} \& {Boynton}}{{Gerend} \&
  {Boynton}}{1976}]{1976ApJ...209..562G}
{Gerend} D.,  {Boynton} P.~E.,  1976, \mn@doi [\apj] {10.1086/154751}, \href
  {https://ui.adsabs.harvard.edu/abs/1976ApJ...209..562G} {209, 562}

\bibitem[\protect\citeauthoryear{{Giacconi}, {Gursky}, {Kellogg}, {Levinson},
  {Schreier}  \& {Tananbaum}}{{Giacconi} et~al.}{1973}]{1973ApJ...184..227G}
{Giacconi} R.,  {Gursky} H.,  {Kellogg} E.,  {Levinson} R.,  {Schreier} E.,
  {Tananbaum} H.,  1973, \mn@doi [\apj] {10.1086/152321}, \href
  {https://ui.adsabs.harvard.edu/abs/1973ApJ...184..227G} {184, 227}

\bibitem[\protect\citeauthoryear{{Jahoda} et~al.,}{{Jahoda}
  et~al.}{2019}]{2019arXiv190710190J}
{Jahoda} K.,  et~al., 2019, arXiv e-prints, \href
  {https://ui.adsabs.harvard.edu/abs/2019arXiv190710190J} {p. arXiv:1907.10190}

\bibitem[\protect\citeauthoryear{{Ji}, {Staubert}, {Ducci}, {Santangelo},
  {Zhang}  \& {Chang}}{{Ji} et~al.}{2019}]{2019MNRAS.484.3797J}
{Ji} L.,  {Staubert} R.,  {Ducci} L.,  {Santangelo} A.,  {Zhang} S.,   {Chang}
  Z.,  2019, \mn@doi [\mnras] {10.1093/mnras/stz264}, \href
  {https://ui.adsabs.harvard.edu/abs/2019MNRAS.484.3797J} {484, 3797}

\bibitem[\protect\citeauthoryear{{Kawashima}, {Mineshige}, {Ohsuga}  \&
  {Ogawa}}{{Kawashima} et~al.}{2016}]{2016PASJ...68...83K}
{Kawashima} T.,  {Mineshige} S.,  {Ohsuga} K.,   {Ogawa} T.,  2016, \mn@doi
  [\pasj] {10.1093/pasj/psw075}, \href
  {https://ui.adsabs.harvard.edu/abs/2016PASJ...68...83K} {68, 83}

\bibitem[\protect\citeauthoryear{{Klein}, {Arons}, {Jernigan}  \&
  {Hsu}}{{Klein} et~al.}{1996}]{1996ApJ...457L..85K}
{Klein} R.~I.,  {Arons} J.,  {Jernigan} G.,   {Hsu} J. J.~L.,  1996, \mn@doi
  [\apj] {10.1086/309897}, \href
  {https://ui.adsabs.harvard.edu/abs/1996ApJ...457L..85K} {457, L85}

\bibitem[\protect\citeauthoryear{{Klochkov} et~al.,}{{Klochkov}
  et~al.}{2008}]{2008A&A...482..907K}
{Klochkov} D.,  et~al., 2008, \mn@doi [\aap] {10.1051/0004-6361:20078953},
  \href {https://ui.adsabs.harvard.edu/abs/2008A&A...482..907K} {482, 907}

\bibitem[\protect\citeauthoryear{{Krawczynski} et~al.,}{{Krawczynski}
  et~al.}{2019}]{2019arXiv190409313K}
{Krawczynski} H.,  et~al., 2019, arXiv e-prints, \href
  {https://ui.adsabs.harvard.edu/abs/2019arXiv190409313K} {p. arXiv:1904.09313}

\bibitem[\protect\citeauthoryear{{Kreykenbohm}, {Wilms}, {Coburn}, {Kuster},
  {Rothschild}, {Heindl}, {Kretschmar}  \& {Staubert}}{{Kreykenbohm}
  et~al.}{2004}]{2004A&A...427..975K}
{Kreykenbohm} I.,  {Wilms} J.,  {Coburn} W.,  {Kuster} M.,  {Rothschild} R.~E.,
   {Heindl} W.~A.,  {Kretschmar} P.,   {Staubert} R.,  2004, \mn@doi [\aap]
  {10.1051/0004-6361:20035836}, \href
  {https://ui.adsabs.harvard.edu/abs/2004A&A...427..975K} {427, 975}

\bibitem[\protect\citeauthoryear{{Leahy}}{{Leahy}}{2002}]{2002MNRAS.334..847L}
{Leahy} D.~A.,  2002, \mn@doi [\mnras] {10.1046/j.1365-8711.2002.05547.x},
  \href {https://ui.adsabs.harvard.edu/abs/2002MNRAS.334..847L} {334, 847}

\bibitem[\protect\citeauthoryear{{Lutovinov}, {Tsygankov}, {Suleimanov},
  {Mushtukov}, {Doroshenko}, {Nagirner}  \& {Poutanen}}{{Lutovinov}
  et~al.}{2015}]{2015MNRAS.448.2175L}
{Lutovinov} A.~A.,  {Tsygankov} S.~S.,  {Suleimanov} V.~F.,  {Mushtukov} A.~A.,
   {Doroshenko} V.,  {Nagirner} D.~I.,   {Poutanen} J.,  2015, \mn@doi [\mnras]
  {10.1093/mnras/stv125}, \href
  {https://ui.adsabs.harvard.edu/abs/2015MNRAS.448.2175L} {448, 2175}

\bibitem[\protect\citeauthoryear{{M{\'e}sz{\'a}ros} \&
  {Nagel}}{{M{\'e}sz{\'a}ros} \& {Nagel}}{1985a}]{1985ApJ...298..147M}
{M{\'e}sz{\'a}ros} P.,  {Nagel} W.,  1985a, \mn@doi [\apj] {10.1086/163594},
  \href {http://adsabs.harvard.edu/abs/1985ApJ...298..147M} {298, 147}

\bibitem[\protect\citeauthoryear{{M{\'e}sz{\'a}ros} \&
  {Nagel}}{{M{\'e}sz{\'a}ros} \& {Nagel}}{1985b}]{1985ApJ...299..138M}
{M{\'e}sz{\'a}ros} P.,  {Nagel} W.,  1985b, \mn@doi [\apj] {10.1086/163687},
  \href {https://ui.adsabs.harvard.edu/abs/1985ApJ...299..138M} {299, 138}

\bibitem[\protect\citeauthoryear{{Mushtukov}, {Tsygankov}, {Serber},
  {Suleimanov}  \& {Poutanen}}{{Mushtukov} et~al.}{2015}]{2015MNRAS.454.2714M}
{Mushtukov} A.~A.,  {Tsygankov} S.~S.,  {Serber} A. e.~V.,  {Suleimanov} V.~F.,
    {Poutanen} J.,  2015, \mn@doi [\mnras] {10.1093/mnras/stv2182}, \href
  {https://ui.adsabs.harvard.edu/abs/2015MNRAS.454.2714M} {454, 2714}

\bibitem[\protect\citeauthoryear{{Mushtukov}, {Verhagen}, {Tsygankov}, {van der
  Klis}, {Lutovinov}  \& {Larchenkova}}{{Mushtukov}
  et~al.}{2018}]{2018MNRAS.474.5425M}
{Mushtukov} A.~A.,  {Verhagen} P.~A.,  {Tsygankov} S.~S.,  {van der Klis} M.,
  {Lutovinov} A.~A.,   {Larchenkova} T.~I.,  2018, \mn@doi [\mnras]
  {10.1093/mnras/stx2905}, \href
  {https://ui.adsabs.harvard.edu/abs/2018MNRAS.474.5425M} {474, 5425}

\bibitem[\protect\citeauthoryear{{Nagel}}{{Nagel}}{1981}]{1981ApJ...251..278N}
{Nagel} W.,  1981, \mn@doi [\apj] {10.1086/159463}, \href
  {https://ui.adsabs.harvard.edu/abs/1981ApJ...251..278N} {251, 278}

\bibitem[\protect\citeauthoryear{{Nishimura}}{{Nishimura}}{2014}]{2014ApJ...781...30N}
{Nishimura} O.,  2014, \mn@doi [\apj] {10.1088/0004-637X/781/1/30}, \href
  {https://ui.adsabs.harvard.edu/abs/2014ApJ...781...30N} {781, 30}

\bibitem[\protect\citeauthoryear{{Postnov}, {Shakura}, {Staubert},
  {Kochetkova}, {Klochkov}  \& {Wilms}}{{Postnov}
  et~al.}{2013}]{2013MNRAS.435.1147P}
{Postnov} K.,  {Shakura} N.,  {Staubert} R.,  {Kochetkova} A.,  {Klochkov} D.,
   {Wilms} J.,  2013, \mn@doi [\mnras] {10.1093/mnras/stt1363}, \href
  {https://ui.adsabs.harvard.edu/abs/2013MNRAS.435.1147P} {435, 1147}

\bibitem[\protect\citeauthoryear{{Poutanen}, {Mushtukov}, {Suleimanov},
  {Tsygankov}, {Nagirner}, {Doroshenko}  \& {Lutovinov}}{{Poutanen}
  et~al.}{2013}]{2013ApJ...777..115P}
{Poutanen} J.,  {Mushtukov} A.~A.,  {Suleimanov} V.~F.,  {Tsygankov} S.~S.,
  {Nagirner} D.~I.,  {Doroshenko} V.,   {Lutovinov} A. e.~A.,  2013, \mn@doi
  [\apj] {10.1088/0004-637X/777/2/115}, \href
  {https://ui.adsabs.harvard.edu/abs/2013ApJ...777..115P} {777, 115}

\bibitem[\protect\citeauthoryear{{Reynolds}, {Quaintrell}, {Still}, {Roche},
  {Chakrabarty}  \& {Levine}}{{Reynolds} et~al.}{1997}]{1997MNRAS.288...43R}
{Reynolds} A.~P.,  {Quaintrell} H.,  {Still} M.~D.,  {Roche} P.,  {Chakrabarty}
  D.,   {Levine} S.~E.,  1997, \mn@doi [\mnras] {10.1093/mnras/288.1.43}, \href
  {https://ui.adsabs.harvard.edu/abs/1997MNRAS.288...43R} {288, 43}

\bibitem[\protect\citeauthoryear{{Schnittman} \& {Krolik}}{{Schnittman} \&
  {Krolik}}{2009}]{2009apj...701.1175s}
{Schnittman} J.~D.,  {Krolik} J.~H.,  2009, \mn@doi [\apj]
  {10.1088/0004-637X/701/2/1175}, \href
  {http://adsabs.harvard.edu/abs/2009ApJ...701.1175S} {701, 1175}

\bibitem[\protect\citeauthoryear{{Schnittman} \& {Krolik}}{{Schnittman} \&
  {Krolik}}{2010}]{2010apj...712..908s}
{Schnittman} J.~D.,  {Krolik} J.~H.,  2010, \mn@doi [\apj]
  {10.1088/0004-637X/712/2/908}, \href
  {http://adsabs.harvard.edu/abs/2010ApJ...712..908S} {712, 908}

\bibitem[\protect\citeauthoryear{{Scott} \& {Leahy}}{{Scott} \&
  {Leahy}}{1999}]{1999ApJ...510..974S}
{Scott} D.~M.,  {Leahy} D.~A.,  1999, \mn@doi [\apj] {10.1086/306631}, \href
  {https://ui.adsabs.harvard.edu/abs/1999ApJ...510..974S} {510, 974}

\bibitem[\protect\citeauthoryear{{Scott}, {Leahy}  \& {Wilson}}{{Scott}
  et~al.}{2000}]{2000ApJ...539..392S}
{Scott} D.~M.,  {Leahy} D.~A.,   {Wilson} R.~B.,  2000, \mn@doi [\apj]
  {10.1086/309203}, \href
  {https://ui.adsabs.harvard.edu/abs/2000ApJ...539..392S} {539, 392}

\bibitem[\protect\citeauthoryear{{She} et~al.,}{{She}
  et~al.}{2015}]{2015SPIE.9601E..0IS}
{She} R.,  et~al., 2015, in UV, X-Ray, and Gamma-Ray Space Instrumentation for
  Astronomy XIX. p. 96010I (\mn@eprint {arXiv} {1509.04392}),
  \mn@doi{10.1117/12.2186133}

\bibitem[\protect\citeauthoryear{{Staubert}, {Klochkov}, {Vasco}, {Postnov},
  {Shakura}, {Wilms}  \& {Rothschild}}{{Staubert}
  et~al.}{2013}]{2013A&A...550A.110S}
{Staubert} R.,  {Klochkov} D.,  {Vasco} D.,  {Postnov} K.,  {Shakura} N.,
  {Wilms} J.,   {Rothschild} R.~E.,  2013, \mn@doi [\aap]
  {10.1051/0004-6361/201220316}, \href
  {https://ui.adsabs.harvard.edu/abs/2013A&A...550A.110S} {550, A110}

\bibitem[\protect\citeauthoryear{{Staubert}, {Klochkov}, {Wilms}, {Postnov},
  {Shakura}, {Rothschild}, {F{\"u}rst}  \& {Harrison}}{{Staubert}
  et~al.}{2014}]{2014A&A...572A.119S}
{Staubert} R.,  {Klochkov} D.,  {Wilms} J.,  {Postnov} K.,  {Shakura} N.~I.,
  {Rothschild} R.~E.,  {F{\"u}rst} F.,   {Harrison} F.~A.,  2014, \mn@doi
  [\aap] {10.1051/0004-6361/201424203}, \href
  {https://ui.adsabs.harvard.edu/abs/2014A&A...572A.119S} {572, A119}

\bibitem[\protect\citeauthoryear{{Tananbaum}, {Gursky}, {Kellogg}, {Levinson},
  {Schreier}  \& {Giacconi}}{{Tananbaum} et~al.}{1972}]{1972ApJ...174L.143T}
{Tananbaum} H.,  {Gursky} H.,  {Kellogg} E.~M.,  {Levinson} R.,  {Schreier} E.,
    {Giacconi} R.,  1972, \mn@doi [\apj] {10.1086/180968}, \href
  {https://ui.adsabs.harvard.edu/abs/1972ApJ...174L.143T} {174, L143}

\bibitem[\protect\citeauthoryear{{Truemper}, {Pietsch}, {Reppin}, {Voges},
  {Staubert}  \& {Kendziorra}}{{Truemper} et~al.}{1978}]{1978ApJ...219L.105T}
{Truemper} J.,  {Pietsch} W.,  {Reppin} C.,  {Voges} W.,  {Staubert} R.,
  {Kendziorra} E.,  1978, \mn@doi [\apjl] {10.1086/182617}, \href
  {https://ui.adsabs.harvard.edu/abs/1978ApJ...219L.105T} {219, L105}

\bibitem[\protect\citeauthoryear{{Truemper}, {Kahabka}, {Oegelman}, {Pietsch}
  \& {Voges}}{{Truemper} et~al.}{1986}]{1986ApJ...300L..63T}
{Truemper} J.,  {Kahabka} P.,  {Oegelman} H.,  {Pietsch} W.,   {Voges} W.,
  1986, \mn@doi [\apjl] {10.1086/184604}, \href
  {https://ui.adsabs.harvard.edu/abs/1986ApJ...300L..63T} {300, L63}

\bibitem[\protect\citeauthoryear{{Tsygankov}, {Lutovinov}, {Churazov}  \&
  {Sunyaev}}{{Tsygankov} et~al.}{2006}]{2006MNRAS.371...19T}
{Tsygankov} S.~S.,  {Lutovinov} A.~A.,  {Churazov} E.~M.,   {Sunyaev} R.~A.,
  2006, \mn@doi [\mnras] {10.1111/j.1365-2966.2006.10610.x}, \href
  {https://ui.adsabs.harvard.edu/abs/2006MNRAS.371...19T} {371, 19}

\bibitem[\protect\citeauthoryear{{Vasco}, {Staubert}, {Klochkov}, {Santangelo},
  {Shakura}  \& {Postnov}}{{Vasco} et~al.}{2013}]{2013A&A...550A.111V}
{Vasco} D.,  {Staubert} R.,  {Klochkov} D.,  {Santangelo} A.,  {Shakura} N.,
  {Postnov} K.,  2013, \mn@doi [\aap] {10.1051/0004-6361/201220181}, \href
  {https://ui.adsabs.harvard.edu/abs/2013A&A...550A.111V} {550, A111}

\bibitem[\protect\citeauthoryear{{Weisskopf} et~al.}{{Weisskopf}
  et~al.}{2016}]{2016SPIE.9905E..17W}
{Weisskopf} M.~C.,  et~al., 2016, in Space Telescopes and Instrumentation 2016:
  Ultraviolet to Gamma Ray. p. 990517, \mn@doi{10.1117/12.2235240}

\bibitem[\protect\citeauthoryear{{West}, {Wolfram}  \& {Becker}}{{West}
  et~al.}{2017a}]{2017ApJ...835..129W}
{West} B.~F.,  {Wolfram} K.~D.,   {Becker} P.~A.,  2017a, \mn@doi [\apj]
  {10.3847/1538-4357/835/2/129}, \href
  {https://ui.adsabs.harvard.edu/abs/2017ApJ...835..129W} {835, 129}

\bibitem[\protect\citeauthoryear{{West}, {Wolfram}  \& {Becker}}{{West}
  et~al.}{2017b}]{2017ApJ...835..130W}
{West} B.~F.,  {Wolfram} K.~D.,   {Becker} P.~A.,  2017b, \mn@doi [\apj]
  {10.3847/1538-4357/835/2/130}, \href
  {https://ui.adsabs.harvard.edu/abs/2017ApJ...835..130W} {835, 130}

\bibitem[\protect\citeauthoryear{{Wolff} et~al.,}{{Wolff}
  et~al.}{2016}]{2016ApJ...831..194W}
{Wolff} M.~T.,  et~al., 2016, \mn@doi [\apj] {10.3847/0004-637X/831/2/194},
  \href {https://ui.adsabs.harvard.edu/abs/2016ApJ...831..194W} {831, 194}

\bibitem[\protect\citeauthoryear{{Yahel}}{{Yahel}}{1980}]{1980ApJ...236..911Y}
{Yahel} R.~Z.,  1980, \mn@doi [\apj] {10.1086/157818}, \href
  {https://ui.adsabs.harvard.edu/abs/1980ApJ...236..911Y} {236, 911}

\bibitem[\protect\citeauthoryear{{Zhang} et~al.}{{Zhang}
  et~al.}{2016}]{2016SPIE.9905E..1QZ}
{Zhang} S.~N.,  et~al., 2016, in Space Telescopes and Instrumentation 2016:
  Ultraviolet to Gamma Ray. p. 99051Q (\mn@eprint {arXiv} {1607.08823}),
  \mn@doi{10.1117/12.2232034}

\makeatother
\end{thebibliography}








\bsp	
\label{lastpage}
\end{document}